\documentclass[prb,twocolumn,showpacs,preprintnumbers,amsmath,amssymb]{revtex4}
\usepackage{graphicx}

\begin{document}

\title{Temperature and Dimensionality Dependences of Optical Absorption Spectra in Mott Insulators}

\author{H. Onodera}
\author{T. Tohyama}
\author{S. Maekawa}
\address{Institute for Materials Research, Tohoku University, Sendai, 980-8577, Japan.}
\date{\today}

\begin{abstract}
We investigate the temperature dependence of optical absorption spectra of one-dimensional (1D) and two-dimensional (2D) Mott insulators by using an effective model in the strong-coupling limit of a half-filed Hubbard model.  In the numerically exact diagonalization calculations on finite-size clusters, we find that in 1D the energy position of the absorption edge is almost independent of temperature, while in 2D the edge position shifts to lower energy with increasing temperature.  The different temperature dependence between 1D and 2D is attributed to the difference of the coupling of the charge and spin degrees of freedom.  The implications of the results on experiments are discussed in terms of the dimensionality dependence.  
\end{abstract}


\maketitle

\section{INTRODUCTION}

The charge gap in Mott insulators is a consequence of a strong electron correlation represented by a large on-site Coulomb interaction.  The correlation induces novel phenomena in terms of the interplay of charge and spin degrees of freedom.~\cite{Maekawa}  In one-dimensional (1D) Mott insulators, two particles created by photoexcitation, i.e., an unoccupied site and a doubly occupied site of electrons, can move inside the system without being disturbed by surrounding spins in the background.  This is a manifestation of a separation of the charge and spin degrees of freedom, called the spin-charge separation.  In two-dimensional (2D) Mott insulators, on the other hand, the two particles are expected not to be free from the spin degree of freedom, because the propagation of a carrier is known to induce a spin cloud around the carrier as a consequence of the misaligned spins along the carrier-hopping paths.  Such an interplay of spin and charge is also one of the main subjects of the study of high-temperature superconductivity.

The nature of the two particles in the photoexcited states of the Mott insulators is obtained by examining the linear susceptibility with respect to the applied electric field, which provides information on the dipole-allowed states with odd parity among the photoexcited states.  In addition to the linear susceptibility, the third-order nonlinear optical susceptibility is useful to detect not only the odd-parity states but also the dipole-forbidden states with even parity.  Recently, large values of the nonlinear optical susceptibility have been reported for 1D Mott insulators of copper oxides and halogen-bridged nickel compounds.~\cite{Kishida} ~\cite{Ogasawara}  Analyses of the susceptibilities have suggested that odd- and even-parity states are nearly degenerate with a large transition dipole moment between them.  Theoretically, the nonlinear susceptibility in 1D Mott insulators has been examined by employing the numerically exact diagonalization technique for small clusters of the Hubbard model at half filling.~\cite{Mizuno}  It has been shown that odd- and even-parity states are almost degenerate in the same energy region and that the degeneracy is due to the spin-charge separation and strong on-site Coulomb interaction.~\cite{Tohyama}

In the 2D Mott insulators of copper oxide such as Sr$_2$CuO$_2$Cl$_2$,  the third-order nonlinear optical susceptibility has been reported to be one order of magnitude smaller than that in 1D.~\cite{Ashida}  Theoretical calculations based on the (extended) Hubbard model have shown such a dimensionality dependence.~\cite{Mizuno,Ashida}  In addition to this, from the numerical investigation of the photoexcited states, it has been found that the edge of the distribution of the even-parity states is located lower in energy than that of the odd-parity states in 2D, in contrast to the almost degenerate behavior in 1D.~\cite{Takahashi}  The origin is attributed to the presence of the exchange interaction between spins, implying the important role of  the spin degree of freedom in the photoexcited states. 

So far the dimensionality dependence of the photoexcited states in the Mott insulators was theoretically examined at zero temperature.  With increasing temperature, magnetic excitations are thermally created in the spin background irrespective of dimensionality.  Therefore, it is not obvious whether the phtoexcited states are dependent on dimensionality even at finite temperatures.   In this study, we numerically examine the temperature dependence of optical absorption spectra in a half-filled Hubbard model with large Coulomb interaction, and clarify its dimensionality dependence.  It is found that there are qualitatively different behaviors between 1D and 2D Mott insulators: in 1D the energy position of the absorption edge is almost independent of temperature as long as the temperature is below the exchange interaction between spins, while in 2D the edge position shifts to lower energy with increasing temperature.  Such a contrasting behavior is caused by the difference of the coupling of charge and spin that persists as long as short-range magnetic interactions are present.   The shift of the edge in 2D is consistent with experimental data in Sr$_2$CuO$_2$Cl$_2$.~\cite{Choi,Lovenich}  The implications of our results to experiments will be discussed in terms of the dimensionality dependence and the effect of phonon scattering.  

The rest of this paper is organized as follows.  We introduce an effective Hamiltonian of the half-filled Hubbard model in the strong-cupling limit, and show outlines of the procedure for caluculating optical absorption spectra in Sec.II.  In Sec.III, caluculated results of the temperature dependence of optical absorption spectra in 1D and 2D sysytems are presents.  The distributions of photoexcited states are discussed in terms of the effect of the spin degree of freedom.  The summary is given in Sec.IV.

\section{MODEL AND METHOD}

The Hubbard model is given by $H_\mathrm{Hub}=H_t+H_U$ with $H_t=-t \sum_{i,\delta,\sigma}(c^{\dagger}_{i,\sigma} c_{i+\delta,\sigma} + \mathrm{H.c.})$ and $H_U=U \sum_{i} n_{i, \uparrow} n_{i, \downarrow}$, where $c^{\dagger}_{i,\sigma}$ is the creation operator of an electron with spin $\sigma$ at site $i$, $ n_{i, \sigma}=c^{\dagger}_{i,\sigma}c_{i,\sigma}$, $i+\delta$ stands for the set of nearest neighbor (NN) sites of $i$, $t$ is the NN hopping integral, and $U$ is the on-site Coulomb interaction.

In the strong coupling limit $(U \gg t)$, there is no doubly occupied site in the initial states of optical absorption at low temperatures $( k_\mathrm{B}T \ll U)$.  (Hereafter, the Boltzmann factor is taken to be $k_\mathrm{B}=1$.) In this case, the initial states are described by the Heisenberg model given by
$H_0=J \sum_{i,\delta} 
( \boldsymbol{S}_{i} \cdot \boldsymbol{S}_{i+\delta} - n_{i} n_{i}/4)$, 
where $\boldsymbol{S}_{i}$ is the spin operator with $S=1/2$
 at site $i$, $n_{i}=n_{i, \uparrow} + n_{i, \downarrow}$, and $J=4t^2/U$.

 The photoexcited states, which are the final states of optical absorption, have both one doubly occupied site and one vacant site. An effective Hamiltonian describing the photoexcited states is obtained by restricting the Hilbert spaces to a subspace with one doubly occupied site. By performing the second order perturbation with respect to the hopping term $H_t$ in the Hubbard Hamiltonian, the effective Hamiltonian is given by
\begin{equation}
H_\mathrm{eff}=\Pi_{1} H_t \Pi_{1}
- \frac{1}{U} \Pi_{1} H_t \Pi_{2} H_t \Pi_{1}
+ \frac{1}{U} \Pi_{1} H_t \Pi_{0} H_t \Pi_{1}
+U
\end{equation}
where $\Pi_{0}$, $\Pi_{1}$, and $\Pi_{2}$ are projection operators onto the Hilbert space with zero, one, and two doubly occupied sites, respectively. In addition to them, we introduce an attractive interaction between a doubly occupied and vacant sites. This interaction is described by 
\begin{equation}
H_V=-V \sum_{i,\delta} 
n_{i,\uparrow} n_{i,\downarrow}
\left( 1- n_{i+\delta,\uparrow} \right)
\left( 1- n_{i+\delta,\downarrow} \right).
\end{equation}

The optical absorption spectrum at temperature $T$ and photon energy $\omega$ is given by
\begin{equation}
\varepsilon_2\left(\omega,\beta\right)=\frac{1}{Z}\left(1-e^{- \beta \omega}\right) \sum_I e^{- \beta E_S} \varepsilon_I
\label{eq:eps}
\end{equation}
with
\begin{equation}
\varepsilon_I=\sum_F \left|\frac{\left\langle F \left| j^x \right| I \right\rangle }{E_F-E_I}\right|^2 \delta \left(\omega-E_F+E_I\right),
\label{ep_I}
\end{equation}
where $\beta=1/T$,  the initial state $\left | I \right \rangle$ is an eigenstate of $H_{0}$ with energy $E_{I}$, and the final state $\left | F \right \rangle$ is an eigenstate of $H_\mathrm{eff}$ with energy $E_{F}$.  The $\varepsilon_I$ represents spectral weight from each initial state $\left | I \right \rangle$.  The energy $E_S$ is measured from the ground-state energy of $H_0$ as $E_S=E_I-E_0$, and $Z=\sum_{I} e^{-\beta E_S}$.  The $j^x$ is the $x$ component of the current operator to second order of $t$. 
It is described by~\cite{Eskes}
\begin{eqnarray}
j^x=\mathrm{i}t\sum_{i,\delta,\sigma} \delta_{x} \tilde{c}^{\dagger}_{i+\delta,\sigma}
\tilde{c}_{i,\sigma}
+\mathrm{i}\frac{t^2}{U}\sum_{i,\delta,\delta^{\prime},\sigma}
\left( \delta_{x}-\delta^{\prime}_{x} \right) \nonumber
\\ \times
\left(
\tilde{c}^{\dagger}_{i+\delta,\sigma}
\tilde{c}^{\dagger}_{i,\bar{\sigma}}
\tilde{c}_{i,\bar{\sigma}}
\tilde{c}_{i+\delta^{\prime},\sigma}
-\tilde{c}^{\dagger}_{i+\delta,\sigma}
\tilde{c}^{\dagger}_{i,\sigma}
\tilde{c}_{i+\delta,\sigma}
\tilde{c}_{i+\delta^{\prime},\sigma}
\right).
\end{eqnarray}
The creation and annihilation operators, $\tilde{c}^{\dagger}_{i,\sigma}$ and $\tilde{c}_{i,\sigma}$, are projected onto the subspace with either zero or one doubly occupied site. The $\delta_{x}$ and $\delta^{\prime}_{x}$ are the $x$ components of vectors connecting NN sites.

\begin{figure}
\begin{center}
\caption{\label{fig1}
The temperature dependence of the optical absorption spectra from $T=0$ to $T=0.8t=2.0J$ in a 16-site 1D chain with $U=10t$ $(J=0.4t)$ and $V=1.5t$.  The delta-functions are broadened by a Lorentzian with a width of $0.05t$.}
\end{center}
\end{figure}

We perform numerical calculations of the optical absorption spectrum for 16-, 20-, and 24-site 1D periodic chains, and $\sqrt{20} \times \sqrt{20} $- and $\sqrt{26} \times \sqrt{26} $- square clusters with periodic boundary conditions. For the 16-site chain, we calculate the optical absorption spectra by using all the initial states $\left| I\right>$ and final states $\left| F\right>$ obtained by performing full diagonalizations of $H_0$ and $H_\mathrm{eff}+H_V$.  For the 20-site clusters, all the initial states $\left| I\right>$ are obtained by diagonalizing $H_0$, but $\varepsilon_I$ for each $\left| I\right>$ is calculated by the Lanczos technique.  The Lanczos technique is also used for obtaining $\left| I\right>$ in the case of the 24- and 26-site clusters.  In this case, we obtain the low-energy eigenstates in the energy region of $0\le E_S \le 4J$ as many as possible.

\section{RESULTS AND DISCUSSIONS}

Figure~\ref{fig1} shows $\varepsilon_2$ for the 16-site chain with $U/t=10$ $(J/t=0.4)$ and $V/t=1.5$.  At $T=0$, there are only three peaks in the energy region of $\omega/t \le 10$ because of finite-size system.  With increasing temperature, the number of peaks increases according to the increase of the initial states contributing to the spectra.  To be emphasized here is that the position of absorption edges at around $6.5t$ is almost unchanged as long as the temperature is less than $\sim2J$.  This feature is independent of the size of the chains. The absence of the shift of the edge position is also seen in analytical calculations~\cite{Gebhard} of the optical conductivity at $T=0$ and high temperatures ($T\gg J$). Note that at very high temperatures ($T\gg t$) the edge shifts to lower energy side accompanied by the broadening the spectrum.

\begin{figure}
\begin{center}
\caption{\label{fig2}
(a) $\varepsilon_I$ in a 16-site 1D chain with $U=10t$ $(J=0.4t)$ and $V=1.5t$ from various initial states $\left| I\right>$ with energy $E_S$ and momentum $k$ in units of $\pi$.  The ground-state energy is set to be $E_S=0$.  (b) The size dependence of absorption-edge positions in a 16-site 1D chains with $J=0.4t$ and $J=0.001t$ $(U=400t)$.  The circle, triangle, and square represent the positions from $\left| I\right>$ with momentum $k=0$, $\pi/2$, and $\pi$, respectively.  The solid and open symbols denote the lowest-energy and next-lowest-energy peaks, respectively,  for each $\left| I\right>$ as shown in (a).  The solid and dotted lines are obtained by fitting them linearly to the symbols for $J=0.4t$ and $J=0.001t$, respectively.}
\end{center}
\end{figure}

In order to see the position of the edge more explicitly, we examine the contribution of each initial state $\left | I \right \rangle$ to the edge.  Figuare~\ref{fig2}(a) shows decomposed spectra $\varepsilon_I$ from several low-energy initial state $\left | I \right \rangle$ that correspond to  two-spinon excitations in the Heisenberg model.  In Fig.\ref{fig2}(a), $J=0.4t$, and $E_S$ and $k$ in each panel represent the energy and the total momentum respectively of the corresponding initial state.  The edge positions are found to be not much dependent on the initial states.  To clarify the positions in the thermodynamic limit, we perform the size scaling of the positions of the lowest-energy and next-lowest-energy peaks for representative three total-momentum spaces ($k=0$, $\pi/2$, and $\pi$) in Fig.~\ref{fig2}(b).  In $N\rightarrow\infty$, $N$ being system size, the lowest-energy peaks (solid symbols) converge to nearly the same position within a range of $0.03t$ irrespective of $J/t$ (solid line: $J/t=0.4$, and dotted line: $J/t=0.001$).  The next-lowest peaks also converge close to the lowest ones.  We do not find any initial states that induce absorption edges lower in energy than that from the ground state.  This is the reason why the absorption-edge position is independent of temperature as seen in Fig.~\ref{fig1}.  

What is physics behind such little initial-state dependence of the edge position?  At first sight, the positions seem to be dependent on the initial states because their energies are different from each other.  However, the edge positions are actually almost independent of the initial states as seen in Fig.~\ref{fig2} (a).  This means that the energy variation among the final states which constitute the absorption edges is almost equivalent to that among the initial states.  This is easily understood if we consider the fact that in the strong coupling limit the wavefunctions in the 1D Hubbard model can be given by the product of the charge  part and the spin part, the latter of which is the wavefunction of the Heisenberg model.~\cite{Ogata}  This leads to the same energy shift in the spin part for both the initial and final states.  In other words, the difference of the initial states as well as that of the final one comes from the different spin states but charge state is the same.  Therefore, we can say that underlying physics behind  no temperature dependence of the edge position is the separation of spin and charge degrees of freedom.   

\begin{figure}
\begin{center}
\caption{\label{fig3}
The temperature dependence of the optical absorption spectra  from $T=0$ to $T=0.8t=2.0J$ in (a) 26-site and (b) 20-site 2D square clusters with $U=10t$ ($J=0.4t$) and $V=0$.  The delta-functions are broadened by a Lorentzian with a width of $0.05t$.}
\end{center}
\end{figure}

\begin{figure}
\begin{center}
\caption{\label{fig4}
$\varepsilon_I$ in (a) 26-site and (b) 20-site 2D square clusters with $U=10t$ ($J=0.4t$) and $V=0$ from various initial states $\left| I\right>$ with energy $E_S$ and momentum $\boldsymbol{k}$ in units of $\pi$.  The delta-functions are broadened by a Lorentzian with a width of $0.05t$.}
\end{center}
\end{figure}

Let us move to 2D systems.  Figure~\ref{fig3} shows $\varepsilon_2$ for the $\sqrt{26}\times\sqrt{26}$- and $\sqrt{20}\times\sqrt{20}$-site clusters with $U/t=10$ $(J/t=0.4)$ and $V/t=0$.  At $T=0$, spectral weights are predominantly concentrated on the absorption-edge region, although the spectral shapes slightly depend on the size of systems.  The states near the edges may be described as excitonic states caused by antiferromagnetic interactions.~\cite{Wrobel,Onodera}  With increasing temperature, spectral weights at absorption edge slightly sift to the lower energy side.  Other clusters such as $4\times 4$ and $\sqrt{18}\times\sqrt{18}$ also show similar temperature dependence (not shown).  This dependence is in contrast with the 1D systems discussed above.  We decompose $\varepsilon_2$ into $\varepsilon_I$'s, which are shown in Fig.~\ref{fig4} for several initial states.  The initial states shown are mainly the states constructing a spin-wave dispersion with $S=1$ in the 2D Heisenberg model.  In contrast to Fig.~\ref{fig2}(a), absorption-edge positions are found to be dependent on the initial states.  Since in 2D the motion of photocarriers created in the final states induces magnetic excitations unlike in 1D, the energy variation among the final states are different from that among the initial states.  This could be the origin of the shift of the edge positions.  Therefore, it may be plausible to say that the coupling of spin and charge causes the temperature dependence of the edge positions.  From Fig.~\ref{fig4}, we can also notice that sharp peaks at the edges (excitonic peaks) emerge depending on the initial states.  By examining spin correlation functions of the initial states, we found that the excitonic peaks are pronounced in the absorptions from the initial states in which antiferromagnetic spin correlation remains to be long-range.  This verifies an picture that the excitonic states are of magnetic origin and magnetically ordered states are necessary to realize them.

Finally let us discuss the implications of the results to optical absorption experiments in Mott insulators.  The temperature dependence of  optical absorption has been reported for a 2D Mott  insulator Sr$_2$CuO$_2$Cl$_2$.~\cite{Choi,Lovenich}  The data show the shift of the edge toward lower-energy side with increasing temperature.  This seems to be consistent with our results in 2D Hubbard model.  However, the shift has been explained by taking into account phonon scattering.~\cite{Lovenich}  Since we have not introduced phonon contributions in our model, we can not say the definite conclusion about the origin of the shift.  From our results, we consider that not only the coupling with phonons but also with magnons~\cite{Choi} may be necessary for precise description of the temperature dependence of the optical absorption.  In 1D, shifts of the edge positions similar to 2D systems have been reported.~\cite{Okamoto}  Since these data are different from our results, coupling with phonon should be examined.

\section{SUMMARY}

In summary, we have examined the temperature dependence of optical absorption spectra of a half-filled 1D and 2D Hubbard models in the strong-coupling limit.  By employing the numerically exact diagonalization calculations on finite-size clusters, we have found that in 1D the energy position of the absorption edge is almost independent of temperature, while in 2D the edge position shifts to lower energy with increasing temperature.  The different temperature dependence between 1D and 2D is attributed to the difference of the coupling of the charge and spin degrees of freedom. The analysis of existing experimental data is left until the effect of the electron-phonon coupling on absorption spectra is examined.

\section*{Acknowledgements}

The authors thank H. Okamoto and H. Kishida for valuable discussions. This work was supported by a Grant-in-Aid for scientific Research from the Ministry of Education, Culture, Sports, Science and Technology of Japan, CREST and NAREGI.  The numerical calculations were performed in the supercomputing facilities in ISSP, University of Tokyo, and IMR, Tohoku University.

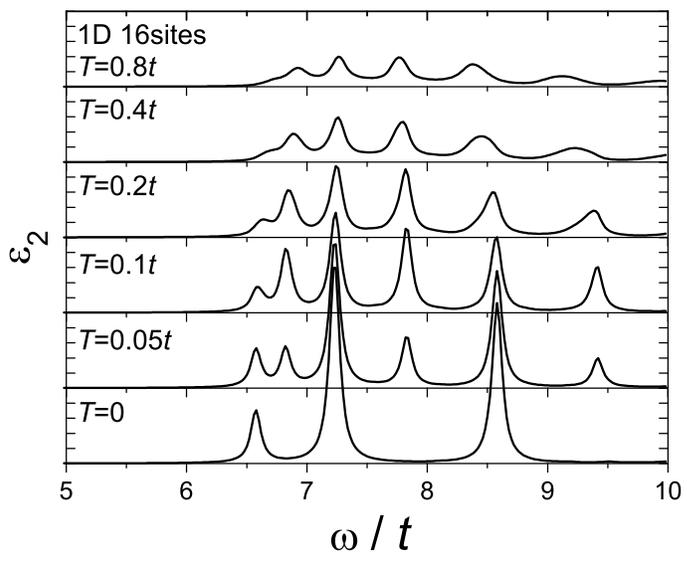
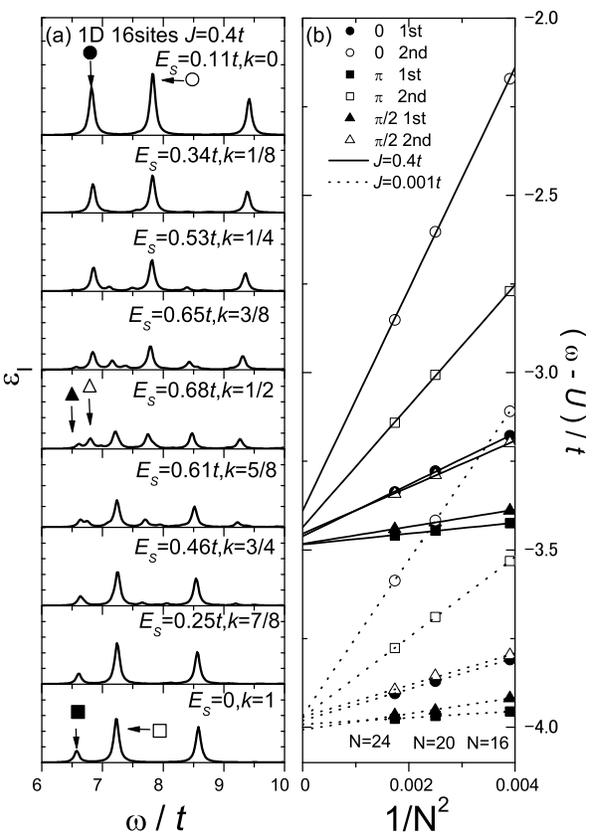
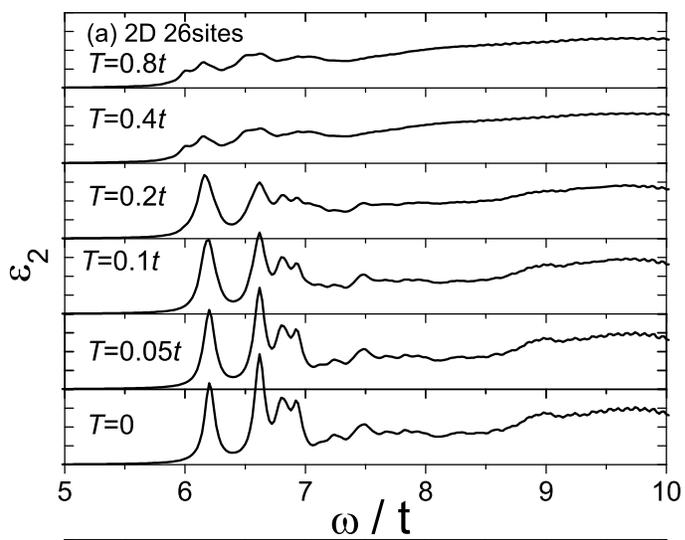
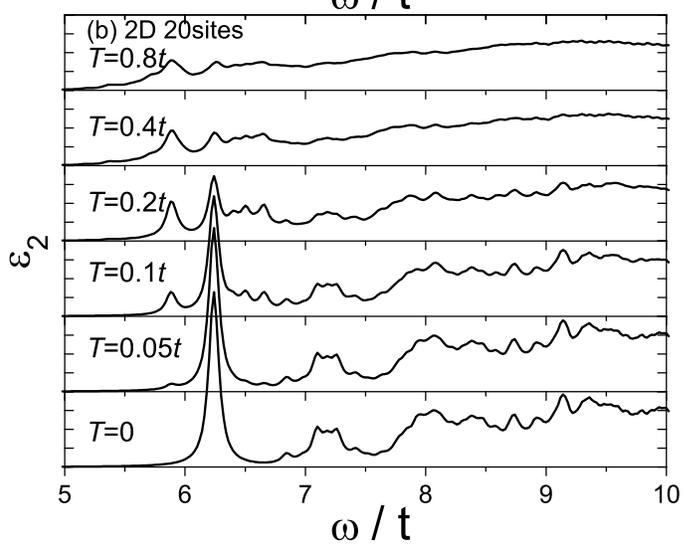
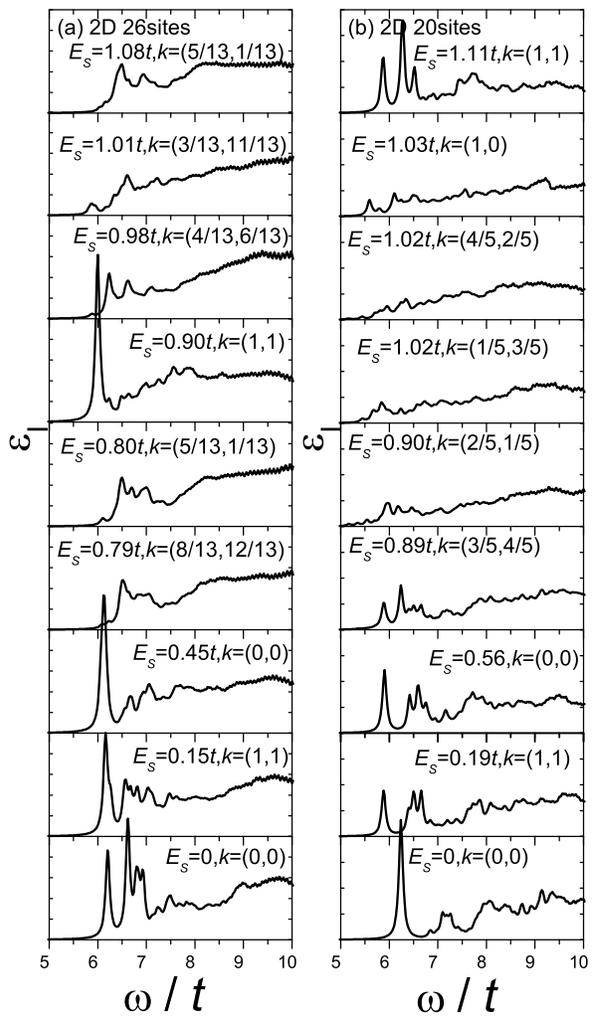
\end{document}